# Version 4.1

# Coronary heart disease, chronic inflammation, and pathogenic social hierarchy: a biological limit to possible reductions in morbidity and mortality

Rodrick Wallace, Ph.D.
The New York State Psychiatric Institute
Deborah Wallace, Ph.D.
Joseph Mailman School of Public Health
Columbia University
Robert G. Wallace, Ph.D.
Dept. of Biology
City University of New York*

November 25, 2003

## Abstract

We suggest that a particular form of social hierarchy, which we characterize as 'pathogenic', can, from the earliest stages of life, exert a formal analog to evolutionary selection pressure, literally writing a permanent developmental image of itself upon immune function as chronic vascular inflammation and its consequences. The staged nature of resulting disease emerges 'naturally' as a rough analog to punctuated equilibrium in evolutionary theory, although selection pressure is a passive filter rather than an active agent like structured psychosocial stress. Exposure differs according to the social constructs of race, class, and ethnicity, accounting in large measure for observed population-level differences in rates of coronary heart disease across industrialized societies. American Apartheid, which enmeshes both majority and minority communities in a social construct of pathogenic hierarchy, appears to present a severe biological limit to continuing declines in coronary heart disease for powerful as well as subordinate subgroups: 'Culture' to use the words of the evolutionary anthropologist Robert Boyd. 'is as much a part of human biology as the enamel on our teeth'.

## Introduction

The origin of 'racial', 'class', and 'ethnic' disparities in health has recently become the center of some debate in the US, with remedies proposed by mainstream authorities characteristically and predictably focused on individual-oriented 'prevention' by altered life-style or related medical 'magic bullet' interventions. See [1] for a comprehensive critique and review. Indeed, for certain US subpopulations, changes in diet, exercise, patterns of smoking and alcohol intake, and so forth, are widely credited with causing markedly declining incidence of coronary heart disease (CHD): the national death rate from CHD for white US males declined from about 420 per 100,000 in 1980 to about 240 by 1997, compared to declines from 350 to 235 for Black males [2]. Declines for both Black and white females have not been as spectacular, starting from a lower 1980 baseline of about 240, and falling to near 150 by 1997.

As [2, 3] note, however, the declines have not been uniformly distributed: CHD mortality rates in the US are especially high in middle-aged black men relative to other race/sex groups. Barnett and Halverson [4] found unexpectedly high rates of premature CHD mortality for African Americans in major metropolitan regions outside the South, despite favorable levels of socioeconomic resources. A recent paper, [5], finds close correlation of CHD mortality with patterns of racial segregation in New York City, one of the world's most segregated urban centers. More generally, [6-9] show that all-cause black-white mortality differences are highest in US metropolitan areas with the greatest racial segregation.

Kiecolt-Glaser et al. [10] discuss how chronic inflammation has recently been linked with a spectrum of conditions associated with aging, including cardiovascular disease (CVD), osteoporosis, arthritis, type II diabetes, certain cancers, and other conditions. The association between CVD and inflammation is mediated by the cytokine IL-6, related to its central role in promoting the production of C-reactive protein (CRP), which [11, 12] describe as an ancient and highly conserved protein secreted by the liver in response to trauma, inflammation, and infection. CRP is a pattern-recognition molecule of the innate immune response keyed to surveillance for altered self and certain pathogens, thus providing early defense and activation of the humoral, adaptive, immune system. It is increasingly seen as a linkage between the two forms of immune response.

*Address correspondence to Rodrick Wallace, PISCS Inc., 549 W 123 St., Suite 16F, New York, NY, 10027. Telephone (212) 865-4766, email rdwall@ix.netcom.com.

Inflammation has recently become central to understanding the etiology of CHD, including its staging as a chronic disease. To paraphrase Blake and Ridker [13], from initial stages of leukocyte recruitment to diseased endothelium, to plaque rupture, inflammatory mechanisms mediate key steps in atherogenesis and its complications. Thus the key to CHD is now seen in the complex basic biology of plaque formation and dynamics rather than in a passive and rather bland lipid storage.

Triggers for inflammation in artherogenesis include hypertension, diabetes, and obesity. Obesity "not only predisposes to insulin resistance and diabetes, but also contributes to atherogenic dyslipidemia...obesity itself promotes inflammation and potentates atherogenesis independent of effects on insulin resistance or lipoproteins" [14].

Libbey et al. [14] conclude that atherothrombosis is more than a disease of lipid accumulation, rather it is a disorder characterized by low-grade vascular inflammation, often associated with traditional risk factors such as central obesity and body mass index. Data implicate inflammatory pathways in all stages of disease, from early atherogenesis, to the progression of lesions, and finally in the thrombotic complications of the disease.

As Ridker [15] states, risk factors for arteriosclerosis and adult-onset diabetes closely overlap and the two disorders may derive from similar antecedents, a mutual inflammatory or genetic basis. Recent studies suggest that baseline levels of IL-6 and CRP which were previously shown to predict onset of atherothrombosis, also predict onset of type II diabetes, even after adjustment for body mass index.

Ridker [15] has the grace to note the limits of such a strict biochemical approach:

> "[T]he clinical hypothesis that an enhanced immune response results in increased plaque vulnerability begs the question as to why a population distribution of inflammation exists in the first place and what the underlying determinants of this distribution might be."

This question is, precisely, the principal focus of our analysis.

Indeed, larger hypotheses are not lacking in the literature, and the newly-recognized cortisol-leptin cycle is worthy of some comment here: Leptin, the 'fat hormone', increases Th1 and suppresses Th2 cytokine production [16] and also stimulates proliferation and activation of circulating monocytes, and may play a direct role in inflammatory processes [17]. Leptin and cortisol have, however, a complex relation. Cortisol, an adrenal stress hormone, and leptin alternate their plasma peaks as part of the normal circadian cycle [18]. Cortisol increases can trigger answer leptin increases [19]. Glucocorticoid levels also influence plasma leptin levels [20]. Thus leptin and the adrenal hormones regulate each other: patterns of stress thus influence weight change, disease resistance, and inflammatory response. Th1/Th2 balance may be heavily influenced, in turn, by the adrenal hormone/leptin balance. Stress imposed on pregnant women may result in changes fetal immune and metabolic processes, with implications for birth weight, fat metabolism and risk for cardiovascular disease and allergenic susceptibility [e.g. 21] over the life course.

These inferences are strengthened by the results of Singhal et al. [22] who found that elevation in leptin was associated with impaired vascular function *independent* of metabolic and inflammatory disturbances associated with obesity.

A long series of articles by Barker and co-workers [23] is consistent with such mechanisms, suggesting that those who develop CHD grow differently from others both in utero and during childhood. Slow growth during fetal life and infancy is followed by accelerated weight gain in childhood, setting a life history trajectory for CHD, type II diabetes, and hypertension. Barker et al. [23] conclude that slow fetal growth might also heighten the body's stress responses and increase vulnerability to poor living conditions in later life. Thus, in his view, CHD is a developmental disorder that originates through two widespread biological phenomena, developmental plasticity and compensatory growth, a speculation consistent with the work of Smith et al. [24] who found that deprivation in childhood influences risk of mortality from CHD in adulthood, although an additive influence of adulthood circumstances is seen in such cases.

These latter results are important for incorporating a second body of research, regarding a particular kind of work stress on the development of CHD. Kivimaki et al. [25] have found that high job strain and 'effort-reward imbalance' increase the risk of cardiovascular mortality, reinforcing a large body of work showing the adverse effect of a 'wage-slavery' work environment on CHD [e.g. 28-30]. Both effort-reward imbalance and lack of control over one's job seem to contribute to development of adult CHD.

Figure 1 redisplays material on hierarchy and health in a recent paper by Singh-Manoux et al. [31]. It is taken from Phase V of the Whitehall II study of London-based office staff, aged 35-55, working in 20 Civil Service departments in 1997 and 1999. The data, covering about 7000 men and 3400 women, show the age-adjusted percent reporting ill-health as a function of self-reported status rank, where 1 is high and 10 is low. Self-reported health is a highly significant predictor of both morbidity and mortality. The results are quite remarkable, both for their evident nonlinearity and for the high prevalence of somaticized distress among the lower-ranked staff. As [1] argues, rank in this case would seem to inextricably convolute lack of job control with significant de-facto deprivation, for example effective exposure to new health information.

Figure 1 is, we claim, very precisely the initial part of a classic 'S-shaped' dose-response curve for exposure of a population to a physiologically-active substance. This one has very nearly reached the 'fifty percent effective concentration level', or EC-50, for the sample, that is, the dosage at which half of the exposed individuals display the observed physiological response, here age-adjusted self-reported illness. The pattern is highly consistent with assertions that social conditions – in this case a particular form of hierarchy – in fact represent 'social exposures' which can be synergistic with other physiologically active agents, for example classic toxic substances. The analysis is, however, complicated by the essential role culture in human life, which, to reiterate the metaphor used by the evolutionary anthropologist Robert Boyd, "is as much a part of human biology as the enamel on our teeth".

CHD seems, then, to be very much a life-history disease as-sociated with a particular kind of sociocultural environment – what we call pathogenic hierarchy. We shall be interested in a model of how such an environment might write itself onto immune function. Our argument, a straightforward adaptation of recent developments in evolutionary theory, couples fundamental human biological mechanisms across multiple scales.

We begin with a restatement of the emerging theory of immune function as immune cognition, and explore linkage with both central nervous system (CNS) cognition, and with the cognitive processes of an embedding 'sociocultural network'. We will then subject this multiply-synergistic cognitive process to patterns of externally-imposed 'structured stress' analogous to evolutionary selection pressure. Using a rate distortion approach, we find such pressure can literally write a distorted version of itself downward in scale onto immune function as chronic inflammation. Thus the special role of culture in human biology [e.g. 31-34], particularly as associated with social hierarchy, becomes directly and organically manifest in the basic biology and dynamics of plaque formation.

That is, for human populations 'cultural factors' like racism, wage slavery, and exaggerated social disparity – what we will call pathogenic social hierarchy – are as much a part of the 'basic biology' of coronary heart disease as are the molecular or biochemical mechanisms of plaque deposition and development.

The theoretical tool we invoke – not the only one possible – is an extension of information theory which, much in the spirit of the Large Deviations Program of applied probability, permits importation of phase transition and related methods from statistical physics, producing chronic, staged, inflammatory disease in a 'highly natural' manner recognizably analogous to 'punctuated equilibrium' in evolutionary process [35-40].

In essence we will argue that pathogenic social hierarchy constitutes a formal, but more directly active, analog to evolutionary selection pressure which, to invert the argument of Ademi et al. [41], literally writes an image of itself upon immune function as chronic vascular inflammation and its sequelae.

### Cognition, immune cognition, and culture

Interactions between the central nervous system (CNS) and the immune system, and between the genetic heritage and the immune system, have become officially recognized and academically codified through journals with titles such as *Neuroimmunology* and *Immunogenetics*. Here we will argue that a cognitive socioculture – a social network embodying culture – in which individuals are embedded, and through which they are both acculturated and function to meet collective challenges of threat and opportunity, may interact strongly with individual immune function to produce a composite entity which might well be labeled an *Immunocultural Condensation* (ICC).

We examine current visions of the interaction between genes and culture, and between the CNS and culture, and follow with a summary of Cohen's view of immune cognition. Next we argue that immune cognition and cognitive socioculture can become fused into a composite entity – the ICC – and that this composite, in turn, can be profoundly influenced by embedding systems of highly structured psychosocial and socioeconomic stressors. In particular, we argue that the internal structure of the external stress – its 'grammar' and 'syntax' – are important in defining the coupling with the ICC.

Wallace [38-40] presents a detailed mathematical model of the ICC and its linkage with structured patterns of psychosocial or socioeconomic stress which is based on adapting renormalization techniques from statistical mechanics to information theory, in the spirit of the Large Deviations Program of applied probability. The necessity of such an approach emerges from examination of the theory of immune cognition.

Increasingly, biologists are roundly excoriating simple genetic reductionism which neglects the role of environment. Lewontin [42], for example, explains that genomes are not 'blueprints,' a favorite public relations metaphor, as genes do not 'encode' for phenotypes. Organisms are instead outgrowths of fluid, conditional interactions between genes and their environments, as well as developmental 'noise.' Organisms, in turn, shape their environments, generating what Lewontin terms a triple helix of cause and effect. Such interpenetration of causal factors may be embodied by an array of organismal phenomena, including, as we shall discuss, culture's relationships with the brain and the immune system. We propose reinterpreting immune function in this light, in particular the coupling of the individual immune system with larger, embedding structures.

The current vision of human biology among evolutionary anthropologists is consistent with Lewontin's analysis and is summarized by Durham [32] as follows:

> "...[G]enes and culture constitute two distinct but interacting systems of inheritance within human populations... [and] information of both kinds has influence, actual or potential, over ... behaviors [which] creates a real and unambiguous symmetry between genes and phenotypes on the one hand, and culture and phenotypes on the other...
>
> [G]enes and culture are best represented as two parallel lines or 'tracks' of hereditary influence on phenotypes..."

With regard to such melding, over hominid evolution genes came to encode for increasing hypersociality, learning, and language skills, so the complex cultural structures which better aid in buffering the local environment became widespread in successful populations [43].

Every successful human population seems to have a core of tool usage, sophisticated language, oral tradition, mythology and music, focused on relatively small family/extended family groupings of various forms. More complex social structures are build on the periphery of this basic genetic/cultural object [35].

At the level of the individual human, the genetic-cultural object appears to be mediated by what evolutionary psychologists postulate are cognitive modules within the human mind [44]. Each module was shaped by natural selection in response to specific environmental and social conundrums Pleistocene hunter-gatherers faced. One set of such domain-specific cognitive adaptations addresses problems of social interchange [45]. The human species' very identity may rest, in part, on its unique evolved capacities for social mediation and cultural transmission. Anthropologist Robert Boyd has remarked that culture is as much a part of human biology as the enamel on our teeth.

Indeed, a brain-and-culture condensation has been adopted as a kind of new orthodoxy in recent studies of human cognition. For example Nisbett et al. [46] review an extensive literature on empirical studies of basic cognitive differences between individuals raised in what they call 'East Asian' and 'Western' cultural heritages. They view Western-based pattern cognition as 'analytic' and East-Asian as 'holistic.' Nisbett et al. [46] find that:

1. Social organization directs attention to some aspects of the perceptual field at the expense of others.

2. What is attended to influences metaphysics.

3. Metaphysics guides tacit epistemology, that is, beliefs about the nature of the world and causality.

4. Epistemology dictates the development and application of some cognitive processes at the expense of others.

5. Social organization can directly affect the plausibility of metaphysical assumptions, such as whether causality should be regarded as residing in the field vs. in the object.

6. Social organization and social practices can directly influence the development and use of cognitive processes such as dialectical vs. logical ones.

Nisbett et al. [46] conclude that tools of thought embody a culture's intellectual history, that tools have theories build into them, and that users accept these theories, albeit unknowingly, when they use these tools.

We may assume, then, the existence of gene-culture and brain-culture condensations.

Recently Atlan and I.R. Cohen [47] have proposed an information-theoretic adaptation of I.R.Cohen's [48, 49] 'cognitive principle' model of immune function and process, a paradigm incorporating pattern recognition behaviors analogous to those of the central nervous system. Their work follows a long tradition of similar 'cognitive' hypotheses regarding immune function, particularly comparison of the immune system's elaborate chemical network with the brain's neural network, an approach which was well expressed in Jerne's 1967 Nobel Prize talk [e.g. 50-56].

Atlan and Cohen [47] describe immune system behaviors of cognitive pattern recognition-and-response as follows:

The meaning of an antigen can be reduced to the type of response the antigen generates. That is, the meaning of an antigen is functionally defined by the response of the immune system. The meaning of an antigen to the system is discernible in the type of immune response produced, not merely whether or not the antigen is perceived by the receptor repertoire. Because the meaning is defined by the type of response there is indeed a response repertoire and not only a receptor repertoire.

To account for immune interpretation Cohen, Grossman, Tauber, Jerne, and many others have proposed a cognitive paradigm for the immune system. According to [47], the immune system can respond to a given antigen in various ways, it has 'options.' Thus the particular response we observe is the outcome of internal processes of weighing and integrating information about the antigen.

In contrast to Burnet's view of the immune response as a simple reflex, it is seen to exercise cognition by the interpolation of a level of information processing between the antigen stimulus and the immune response. A cognitive immune system organizes the information borne by the antigen stimulus within a given context and creates a format suitable for internal processing; the antigen and its context are transcribed internally into the 'chemical language' of the immune system.

Atlan and Cohen's formulation of the cognitive paradigm suggests a language metaphor to describe immune communication by a string of chemical signals. This metaphor is apt because the human and immune languages can be seen to manifest several similarities such as syntax and abstraction. Syntax, for example, enhances both linguistic and immune meaning.

Although individual words and even letters can have their own meanings, an unconnected subject or an unconnected predicate will tend to mean less than does the sentence generated by their connection.

The immune system, in Atlan and Cohen's view, creates a 'language' by linking two ontogenetically different classes of molecules in a syntactical fashion. One class of molecules are the T and B cell receptors for antigens. These molecules are not inherited, but are somatically generated in each individual. The other class of molecules responsible for internal information processing is encoded in the individual's germline.

Meaning, the chosen type of immune response, is the outcome of the concrete connection between the antigen subject and the germline predicate signals.

The transcription of the antigens into processed peptides embedded in a context of germline ancillary signals constitutes the functional 'language' of the immune system. Despite the logic of clonal selection, the immune system does not respond to antigens as they are, but to abstractions of antigens-in-context.

As shown at length in the highly mathematical developments of [37, 38, 40], it is possible to give Atlan and Cohen's language metaphor of meaning-from-response-in-context a precise information-theoretic characterization. In essence, 'choice-in-context' determines a 'dual information source' [37, 38, 40]. It is further possible to place that characterization within the realm of recent developments which propose the 'coevolutionary' mutual entrainment – in a large sense – of different information sources to create larger metalanguages containing the original as subdialects [35-40, 57]. This work also permits treating gene-culture and brain-culture condensations using a similar, unified, conceptual framework of information source 'coevolutionary condensation'. The Atlan and Cohen version of the immune cognition model suggests, then, the possibility that human culture and the human immune system may be jointly convoluted: To 'neuroimmunology' and 'immunogenetics' we add 'immunocultural condensation.'

The evolutionary anthropologists' vision of the world, as we have interpreted it, sees language, culture, gene pool, and individual CNS and immune cognition as intrinsically melded and synergistic. We propose, then, that culture, as embodied in a local cognitive sociocultural network, and individual immune cognition may become a joint entity whose observation may be 'confounded' – and even perhaps masked – by the distinct population genetics associated with assortive mating due to linguistic and cultural isolation.

## Punctuated interpenetration: how the cognitive condensation adapts to pathogenic hierarchy

Ademi et al. [41] see genomic complexity as the amount of information a gene sequence stores about its environment. Something similar can be said of a reverse process: environmental complexity is the amount of information organisms introduce into the environment as a result of their collective actions and interactions [42]. From that interactive perspective Wallace [39, 40] has invoked an information theory formalism, imposing invariance under renormalization on the mutual information characterizing the Rate Distortion Theorem as applied to Ademi's mapping. The result is a description of how a structured environment, through adaptation, literally writes a (necessarily) distorted image of itself onto the genetic structure of an organism in a punctuated manner, while itself often (but not always) being affected by changes in the organism. Wallace and Wallace [36, 37] use punctuated splittings and coagulations of 'languages-on-networks' to represent, respectively, speciation and coevolution.

Once immune cognition and CNS cognition are seen as linked, it is possible to embed the dual information sources associated with those cognitive processes [38] within a matrix defined by a local, but larger and encompassing, cognitive sociocultural network, using an extension of the Rate Distortion Theorem known as Network Information Theory: The three information source 'languages' of the layers of cognitive process interact to create a more complicated mutual information upon which we impose invariance under renormalization to obtain an analog to punctuated equilibrium evolutionary process [37-40]. Using the fundamental asymptotic limit theorems of information theory, we thus examine the punctuated behavior of single, double, and, finally, triple, sets of 'paths' constituting output strings of individual or interacting 'languages', in a large sense.

Note that in evolutionary process the interaction is largely, but not necessarily entirely, throught the passive filter of selection. Pathogenic social hierarchy, as we view it, is a more active instrument of interaction between physiological and psychosocial 'languages'.

Through the particular influence of the local sociocultural network on humans, culture then becomes, quite literally, 'as much a part of human biology as the enamel on our teeth', to extend the metaphor of the evolutionary anthropologist Robert Boyd.

We propose one more iteration, using the Joint Asymptotic Equipartition or Rate Distortion Theorems, which apply to dual interacting information sources. We suppose that the tripartite mutual information representing the interpenetrative coagulation of immune, CNS, and locally 'social' cognition, is itself subjected to a 'selection pressure', i.e. influence by a larger embedding, but highly structured, process representing the power relations between groups. Most typically, these would constitute pathogenic hierarchical systems of imposed economic inequality and deprivation, the historic social construct of racism, patterns of wage-slavery or, very likely, a coherent amalgam of them all. Wallace et al [40] give a full mathematical treatment of such multiple interacting information sources in terms of 'network information theory'.

The result of this iteration, using a Rate Distortion argument [40], is to find that, in Ademi's [41] sense, pathologies of social hierarchy are formally analogous to evolutionary selection pressure, and, depending in a punctuated manner on the degree of coupling between an individual and the embedding context, can literally write a distorted image of themselves down the chain of human biological interpenetration onto the development and functioning of the immune system at every stage of life.

We thus propose that chronic vascular inflammation resulting in coronary heart disease is not merely the passive result of changes in human diet and activity in historical times [e.g. 15], but represents the image of literally inhuman 'racial' and socioeconomic policies, practices, history, and related mechanisms of pathogenic social hierarchy imposed upon the immune system, beginning in utero, and continuing throughout the life course.

Our interpretation is consistent with, but extends slightly, already huge and rapidly growing animal model and 'health disparities' literatures, [e.g. 28-30, 58-63]. Kaplan et al. [61], for example, found that female primates fed an atherogenic diet were markedly graded on risk of CHD inversely according to social status, in spite of the supposed protective effect of female hormones.

One particular consequence of our importation of 'punctuated equilibrium' formalism from evolutionary theory [39, 40] is that the writing of pathogenic social hierarchy onto the human organism through vascular inflammation should itself be recognizably punctuated, accounting in a 'natural' manner for the staged progress of the disease.

The formal analog between pathogenic social hierarchy and evolutionary selection pressure is through our invocation of a punctuated version of the Rate Distortion Theorem. Although formally analogous from that limited mathematical perspective, they are, nonetheless, fundamentally different. Evolutionary selection pressure acts as a passive filter for the survival of mutations in order to write an image of the environment onto genetic structure. Mutator mechanisms of 'second order selection', to use the terminology of Tenallion et al. [64], are a proposed 'semi-active' means by which environmental stressors increase the rate of certain kinds of mutations, but again selection acts as a fitness-filter for survivors. Pathogenic social hierarchy, as we have described it, is not passive, but is more actively convoluted with human biology over the life course. A particular utility of our 'weak' treatment of languages-on-networks, using formalism in the spirit of the Large Deviations Program of applied probability, is that it permits a punctuated equilibrium approach to interaction across such diverse phenomena.

## Discussion: implications for intervention

Our analysis suggests that, under conditions of racism, wage slavery, draconian socioeconomic inequality, and outright material deprivation, an aspirin a day (or some chemical equivalent) will not keep death at bay. That is, pathogenic social hierarchy is a protean and determinedly plieotropic force, having many possible pathways for its biological expression: if not heart disease, then high blood pressure, if not high blood pressure, then cancer, if not cancer, diabetes, if not diabetes, then behavioral pathologies leading to raised rates of violence or substance abuse, and so on. We have explored a particular mechanism by which pathogenic social hierarchy imposes an image of itself on the human immune system through vascular inflammation. Work like that of [9, 65], or [31], as shown in figure 1, implies, however, the existence multiple, competing, pathways along which deprivation, inequality, and injustice operate. These not only write themselves onto molecular mechanisms of 'basic' human biology, but become, as a result of the particular role of culture among humans, literally a part of that basic biology.

The nature of human life in community, and the special role of culture in that life, ensures that individual psychoneuroimmunology cannot be disentangled from social process, its cultural determinants, and their historic trajectory. Psychosocial stress is not some undifferentiated quantity like the pressure under water, but has a complex and coherent cultural grammar and syntax which write themselves as a particular distorted image of pathogenic social hierarchy within the human immune system: chronic vascular inflammation.

For marginalized populations, this is not a simple process amenable to magic bullet interventions. Substance abuse and overeating become mechanisms for self medication and the leavening of distorted leptin/cortisol cycles. Activity and exercise patterns may be constrained by social pathologies representing larger-scale written images of racism [66].

Culture, as a kind of extended generalized 'language', is path-dependent: Changes are almost always based on, and consistent with, preexisting structures, i.e. the burdens of history. A cultural history of pathogenic social hierarchy, then, may continue to write itself on human immune cognition as chronic vascular inflammation, requiring large-scale and very disruptive 'affirmative action' interventions for redress. That is, elaborate, ecosystem-based programs of 'Comprehensive Inflammation Management' (CIM), much like Comprehensive Pest Management in agriculture, may be required. The history of fighting outbreaks of agricultural pests with 'magic bullet' pesticide application is rife with failure, as the organisms simply evolve chemical resistance. The writing of pathogenic social hierarchy onto human immune function over the life course seems to be a fundamental, and likely very plastic, biological mechanism equally unlikely to respond, in the long run, to magic bullet interventions. Rather, an extension of the comprehensive reforms which largely ended the scourge of infectious disease in the late 19th and early 20th centuries seems prerequisite to significant intervention against coronary heart disease and related disorders for marginalized populations within modern industrialized societies.

This analysis has obvious implications for the continued decline of CHD within the US majority population. Our own studies show clearly that the public health impacts of recent massive deindustrialization and deurbanization in the US have not been confined to urbanized minority or working-class communities where they have been focused, but have become 'regionalized' in a very precise sense so as to entrain surrounding suburban counties into both national patterns of hierarchical, and metropolitan regional patterns of spatially contagious, diffusion of emerging infection and behavioral pathology [67-69]. In essence, social disintegration has diffused outward from decaying urban centers, carrying with it both disease and disorder [70]. To use a phrase first coined by Greg Pappas, "concentration is not containment", and the system of American Apartheid [71] has quite simply been unable to limit health impacts to minority communities, a reality starkly contrary to very deeply held and emotionally compelling cultural beliefs.

In precisely the same sense, it seems virtually inevitable that American Apartheid, as expressed in patterns of pathogenic hierarchy entraining all subpopulations, will similarly constitute a very real biological limit, Robert Boyd's sense, to possible declines in CHD among both white and Black subpopulations. Figure 1 suggests that nobody is more enmeshed in, and hence susceptible to, the pathologies of hierarchy than those of a majority whose fundamental cultural assumptions include the social reality of divisions by class and race.

There is some empirical support for this perspective as it affects chronic inflammatory disease: Wallace and Wallace [72] examined the spatial structure of diabetes mortality incidence, a correlate of CHD, in several US metropolitan regions, contrasting the time periods 1979-85 and 1986-94 at the county level. While the overall structure of diabetes mortality was poverty-driven, the New York metropolitan region, one of the most virulently segregated in the US [71], showed a startling *decline* in the strength of the relation between diabetes mortality rate and poverty rate over the two time periods, from $R^2 = 0.44, P = 0.0003$ to $R^2 = 0.16, P = 0.03$. Wallace and Wallace [72] conclude that the marked weakening of the relation for the New York metro region is not a sign of improvement in the lot of the poor, rather it means that high incidence is spilling over into areas with low-to-moderate poverty rates, i.e. high incidence is crossing class lines. The explanation, they infer, may lie in either or both of two hypotheses: the level of stress once associated with poverty is affecting those above the poverty line in this metro region, or the response to stress once concentrated in the population below the poverty line has been adopted by those not living in poverty. Because of the great increase in the proportion of the US population which qualifies as obese, they believe that the explanation is a combination. The stress on the blue collar and white collar classes may lead them to seek relief, with that relief partly in excess food and passive pastimes.

We conclude that American Apartheid, and similar systems of pathogenic social hierarchy, are classic double-edged swords which wound both dominant and subordinate communities, placing a very real biological limit to the possible decline of coronary heart disease. Programs of social and cultural reform affecting marginalized populations will inevitably entrain the powerful as well, to the benefit of all.

### Acknowledgements

This work benefited from support under NIEHS Grant I-P50-ES09600-05, and from earlier monies provided under an Investigator Award in Health Policy Research from the Robert Wood Johnson Foundation.

### References

[1] Link B. and J. Phelan, 2000, Evaluating the fundamental cause explanation for social disparities in health, in *Handbook of Medical Sociology*, Fifth ed., C. Bird, P. Conrad, and A. Fremont (eds.), Prentice-Hall, New Jersey.

[2] Cooper R., et al. (14 more authors), 2000, Trends and disparities in coronary heart disease stroke, and other cardiovascular diseases in the United States: findings of the National Conference on Cardiovascular Disease, *Circulation*, **102**, 3137-3147.

[3] Cooper R., 2001, Social inequality, ethnicity and cardiovascular disease, *International Journal of Epidemiology*, **30**, Supp. 1, S48-52.

[4] Barnett E. and J Halverson, 2000, Disparities in premature coronary heart disease mortality by region and urbanicity among black and white adults ages 35-64, 1985-1995, *Public Health Reports*, **115**, 52-64.

[5] Fang J., S. Madhavan, W. Bosworth and M. Alderman, 1998, Residential segregation and mortality in New York City, *Social Science and Medicine*, **47**, 469-476.

[6] Polednak A., 1991, Black-white differences in infant mortality in 38 SMSA, *American Journal of Public Health*, **81**, 1480-1482.

[7] Polednak A., 1993, Poverty, residential segregation, and black/white mortality ratios in urban areas, *Journal of Health Care for the Poor and Underserviced*, **4**, 363-373.

[8] Polednak A., 1996, Segregation, discrimination and mortality in US Blacks, *Ethnicity and Disease*, **6**, 99-108.

[9] Collins C. and D. Williams, 1996, Examining the black-white adult mortality: the role of residential segregation, for 25th Public Health Conf. on Records and Statistics and the National Committee on Vital and Health Statistics 45th Anniversary Symp., July 17-19, Washington DC.

[10] Kiecolt-Glaser J., L. McGuier, T. Robles, and R. Glaser, 2002, Emotions, morbidity, and mortality: new perspectives from psychoneuroimmunology, *Annual Review of Psychology*, **53**, 83-107.

[11] Du Clos T., 2000, Function of C-reactive protein, *Annals of Medicine*, **32**, 274-278.

[12] Volanakis J., 2001, Human C-reactive protein: expression, structure, and function, *Molecular Immunity*, **38**, 189-197.

[13] Blake G. and P. Ridker, 2001, Inflammatory mechanisms in atherosclerosis: from laboratory evidence to clinical application, *Italian Heart Journal*, **2**, 796-800. Blake G. and P. Ridker, 2001, Inflammatory bio-markers and cardiovascular risk prediction, *Journal of Internal Medicine*, **252**, 283-294.

[14] Libbey P., P. Ridker, and A. Maseri, 2002, Inflammation and atherosclorosis, *Circulation*, **105**, 1135-1143.

[15] Ridker P., 2002, On evolutionary biology, inflammation, infection, and the causes of atherosclerosis, *Circulation*, **105**, 2-4.

[16] Lord G., G. Matarese, J. Howard, R. Baker, S. Bloom and R. Lechler, 1998, Leptin modulates the T-cell immune response and reverses starvation-induced immunosuppression, *Nature*, **394**, 897-901.

[17] Santos-Alvarez J., R. Governa and V. Sanchez-Margalet, 1999, Human leptin stimulates proliferation and activation of circulating monocytes, *Cellular Immunology*, **194**, 6-11.

[18] Bornstein S., J. Licinio, R. Tauchnitz, L. Engelmann, A. Negrao, P. Gould and G. Chrousos, 1998, Plasma leptin levels are increased in survivors of acute sepsis: associated loss of diurnal rhythm, in cortisol and leptin secretion, *Journal of Clinical Endicrinology and Metabolism*, **83**, 280-283.

[19] Newcomer J., G. Selke, H. Melson, J. Gross, G. Vogler and S. Dagogo-Jack, 1988, Dose-dependent cortisol-induced increases in plasma leptin concentration in healthy humans, *Archives of General Psychiatry*, **55**, 995-1000.

[20] Eliman A., U. Knutsson, M. Gronnegard, A. Steirna, K. Albertsson-Wiklan and C. Marcus, 1998, Variations in glucocorticoid levels within the physiological range affect plasma leptin levels, *European Journal of Endocrinology*, **139**, 615-620.

[21] Wallace R., Wallace D., and M. Fullilove, 2003, Community lynching and the US asthma epidemic. Submitted.

[22] Singhal A., I. Farooque, T. Cole, S. O'Rahilly, M. Fewtress, M. Kattenhorn, A. Lucas, and J. Deanfield, 2002, Influence of leptin on arterial distensibility: a novel link between obesity and cardiovascular disease?, *Circulation*, **15**, 1919-1924.

[23] Barker D., 2002, Fetal programming of coronary heart disease, *Trends in Endocrinology and Metabolism*, **13**, 364, Barker D. T. Forsen, A. Uutela, C. Osmond, and J. Eriksson, 2002, Size at birth and resilience to effects of poor living conditions in adult life: longitudinal study, *British Medical Journal*, **323**, 1261-1262, Osmond C. and D. Barker, 2000, *Environmental Health Perspectives*, **108**, Suppl. 3, 545-553, Godfrey K. and D. Barker, 2001, Fetal programming and adult health, *Public Health and Nutrition*, **4**, 611-624.

[24] Smith G., C. Hart, D. Blane, and D. Hole, 1998, Adverse socioeconomic conditions in childhood and cause specific adult mortality: prospective observational study, *British Medical Journal*, **317**, 1631-1635.

[25] Kivimaki M., P. Leino-Arjas, R. Luukkonen, H. Riihimaki, J. Vahtera, and J. Kirjonen, 2002, Work stress and risk of cardiovascular mortality: prospective cohort study of industrial employees, *British Medical Journal*, **325**, xxx-xxx.

[27] Marmot M., H. Bosma, H. Hemingway, E. Brunner, and S. Stansfeld, 1997, Contribution of job control and other risk factors to social variations in coronary heart disease incidence, *The Lancet*, **350**, 235-239.

[28] Bosma H., M. Marmot, H. Hemingway, A. Nicholson, E. Brunner, and S. Stansfeld, 1997, Low job control and risk of coronary heart disease in Whitehall II (prospective cohort) study, *British Medical Journal*, **314** 558-565.

[29] Bosma H., S. Stansfeld, and M. Marmot, 1998, Job control, personal characteristics, and heart disease, *Journal of Occupational Health and Psychology*, **3**, 402-409.

[30] Bosma H., R. Peter, J. Siegrist, and M. Marmot, 2001, Two alternative job stress models and the risk of coronary heart disease, *American Journal of Public Health*, **88**, 68-74.

[31] Singh-Manoux A., N. Adler, and M. Marmot, 2003, Subjective social status: its determinants and its association with measures of ill-health in the Whitehall II study, *Social Science and Medicine*, 56:1321-1333.

[32] Durham W., 1991, *Coevolution: Genes, Culture and Human Diversity*, Stanford University Press, Palo Alto, CA.

[33] Richerson P. and R. Boyd, 1995, "The evolution of human hypersociality." Paper for Ringberg Castle Symposium on Ideology, Warfare and Indoctrinability (January, 1995), and for HBES meeting, 1995.

[34] Richerson P. and R. Boyd, 1998, "Complex societies: the evolutionary origins of a crude superorganism," to appear.

[35] Wallace R., and R.G. Wallace, 1998, Information theory, scaling laws and the thermodynamics of evolution, *Journal of Theoretical Biology*, **192**, 545-559.

[36] Wallace R., and R.G. Wallace, 1999, Organisms, organizations and interactions: an information theory approach to biocultural evolution, *BioSystems*, **51**, 101-119.

[37] Wallace R., 2000, Language and coherent neural amplification in hierarchical systems: renormalization and the dual information source of a generalized spatiotemporal stochastic resonance, *International Journal of Bifurcation and Chaos*, **10**, 493-502.

[38] Wallace R., 2002, Immune cognition and vaccine strategy: pathogenic challenge and ecological resilience, *Open Systems and Information Dynamics*, **9**, 1-30.

[39] Wallace R., 2002, Adaptation, punctuation and information, a rate-distortion approach to non-cognitive 'learning plateaus' in evolutionary process, *Acta Biotheoretica*, **50**, 101-116.

[40] Wallace R., D. Wallace, and R.G. Wallace, 2003, Toward cultural oncology: the evolutionary information dynamics of cancer, *Open Systems and Information Dynamics*, 10(2):xxx-xxx.

[41] Ademi C., C. Ofria and T. Collier, 2000, Evolution of biological complexity. *Proceedings of the National Academy of Sciences*, **97**, 4463-4468.

[42] Lewontin R., 2000, *The Triple Helix: Gene, Organism and Environment*, Harvard University Press, Cambridge, MA.

[43] Bonner J., 1980, *The Evolution of Culture in Animals*, Princeton University Press, Princeton, NJ.

[44] Barkow J., L. Cosmides and J. Tooby (eds.), 1992, *The Adapted Mind: Biological approaches to mind and culture*, University of Toronto Press.

[45] Cosmides L. and J. Tooby, 1992, Cognitive adaptations for social exchange, in *The Adapted Mind: Evolutionary Psychology and the Generation of Culture*, Oxford University Press, New York.

[46] Nisbett R., K. Peng, C. Incheol and A. Norenzayan, 2001, Culture and systems of thought: holistic vs. analytic cognition, *Psychological Review*, **108**, 291-310.

[47] Atlan H. and I.R.Cohen, 1998, Immune information, self-organization and meaning, *International Immunology*, **10**, 711-717.

[48] Cohen I.R., 1992, The cognitive principle challenges clonal selection, *Immunology Today*, **13**, 441-444.

[49] Cohen I.R., 2000, *Tending Adam's Garden: evolving the cognitive immune self*, Academic Press, New York.

[50] Grossman Z., 1989, The concept of idiotypic network: deficient or premature? In: H. Atlan and IR Cohen, (eds.), *Theories of Immune Networks*, Springer Verlag, Berlin, p. 3852.

[51] Grossman Z., 1992a, Contextual discrimination of antigens by the immune system: towards a unifying hypothesis, in: A. Perelson and G. Weisbch, (eds.) *Theoretical and Experimental Insights into Immunology*, Springer Verlag, p. 7189.

[52] Grossman Z., 1992b, *International Journal of Neuroscience*, 64:275.

[53] Grossman Z., 1993, Cellular tolerance as a dynamic state of the adaptable lymphocyte, *Immunology Reviews*, 133:45-73.

[54] Grossman Z., 2000, Round 3, *Seminars in Immunology*, 12:313-318.

[55] Tauber A., 1998, Conceptual shifts in immunology: Comments on the 'two-way paradigm'. In K. Schaffner and T. Starzl (eds.), Paradigm Changes in Organ Transplantation, *Theoretical Medicine and Bioethics*, 19:457-473.

[56] Podolsky S. and A. Tauber, 1997, *The generation of diversity: Clonal selection theory and the rise of molecular biology*, Harvard University Press.

[57] Wallace R. and R. Fullilove, 1999, Why simple regression models work so well describing 'risk behaviors' in the USA, *Environment and Planning A*, **31**, 719-734.

[58] Schapiro S., P. Nebete, J. Perlman, M. Bloomsmith, and K. Sastry, 1998, Effects of dominance status and environmental enrichment on cell-mediated immunity in rhesus macaques, *Applied Animal Behavioral Science*, **56**, 319-332.

[59] De Groot J., M. Ruis, J. Scholten, J. Koolhass, and W. Boersma, 2001, Long-term effects of social stress on antiviral immunity in pigs, *Physiology and Behavior*, **73**, 145-158.

[60] Gryazeva A., I. Shurlygina, L. Verbitskaya, E. Mel'nikova, N. Kudryavtseva, and V. Trufakin, 2001, Changes in various measures of immune status in mice subject to social conflict, *Neuroscience and Behavioral Physiology*, **31**, 75-81.

[61] Kaplan J., M. Adams, T. Clarkson, S. Manuck, C. Shively, and J. Williams, 1996, Psychosocial factors, sex differences, and atherosclerosis: lessons from animal models, *Psychosomatic Medicine*, **58**, 598-611.

[62] Wilkinson, R., 1996, *Unhealthy Societies: The afflictions of inequality*, Routledge, London.

[63] Karlsen S. and J. Nazroo, 2002, Relation between racial discrimination, social class, and health among ethnic minority groups, *American Journal of Public Health*, **92**, 624-631.

[64] Tenallion O., F. Taddei, M. Radman, and I. Matic, 2001, Second-order selection in bacterial evolution: selection acting on mutation and recombination rates in the course of adaptation, *Research on Microbiology*, **115**, 11-16.

[65] McCord C. and H. Freeman, 1990, Excess mortality in Harlem, *New England Journal of Medicine*, **322**, 173-183.

[66] Wallace R., M. Fullilove, and A. Flisher, 1996, AIDS, violence and behavioral coding: information theory, risk behavior and dynamic process on core-group sociogeographic networks, *Social Science and Medicine*, **43**, 339-352.

[67] Wallace R., and D. Wallace, 1997, Community marginalization and the diffusion of disease and disorder in the United States, *British Medical Journal*, **314**, 1341-1345.

[68] Wallace R., and D. Wallace, 1999, Emerging infections and nested martingales: the entrainment of affluent populations into the disease ecology of marginalization, *Environment and Planning A*, **31**, 1787-1803.

[69] Wallace D., and R. Wallace, 1999, *A Plague on Your Houses*, Verso, New York.

[70] Wallace R., D. Wallace, and H. Andrews, 1997, AIDS, tuberculosis, violent crime and low birthweight in eight US metropolitan areas: stochastic resonance and the diffusion of inner-city markers, *Environment and Planning A*, **29**, 525-555.

[71] Massey D., and N. Denton, 1993, *American Apartheid: Segregation and the Making of the Underclass*, Harvard University Press, Cambridge, MA.

[72] Wallace D., and R. Wallace, 1998, Geography of asthma and diabetes over eight US metropolitan regions, *Journal of Environmental Disease and Health Care Planning*, **3**, 73-88.

Figure Caption

**Figure 1.** Redisplay of data from [31]: Dose-response curve of age-adjusted prevalence of self-reported ill-health vs. self-reported status rank for men and women. 1 = high status, 10 = low status. Note that the upper point is very near the 'EC-50' level in this population. Self-reported health is a highly significant predictor of both morbidity and mortality.

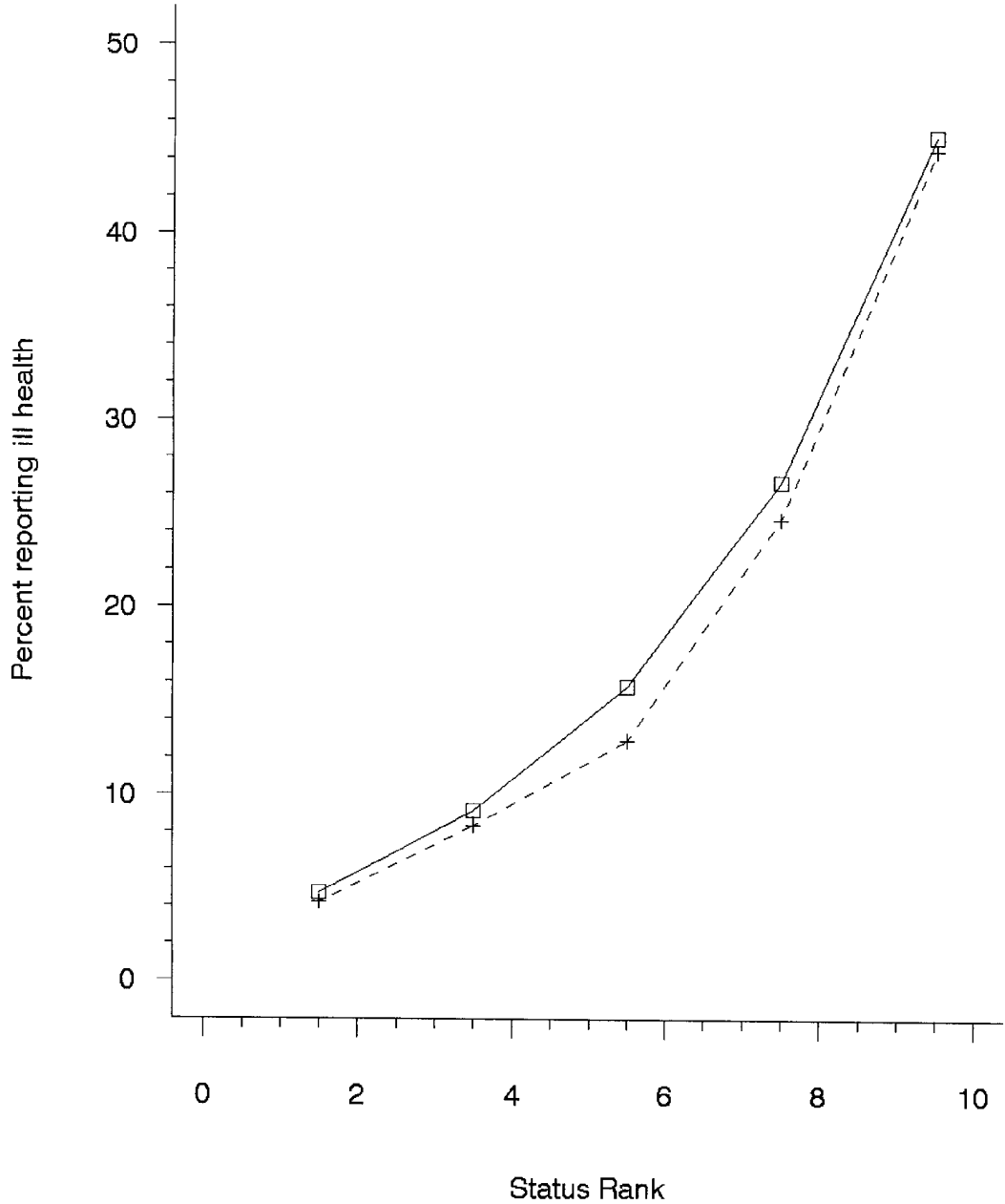